\pdfoutput=1
\documentclass[a4paper,10pt]{article}
\usepackage{jheppub,amsmath,amsfonts,amssymb,xcolor,graphicx,hyperref}
\usepackage[utf8x]{inputenc}
\usepackage[section]{placeins}
\usepackage{cancel}

\usepackage{subfigure,color,graphicx,hyperref,dsfont,slashed,multirow}
\usepackage{fontenc}
\usepackage{pbox}
\usepackage{cancel}
\usepackage{setspace}
\usepackage{float}
\usepackage{mathtools,nccmath}%
\usepackage{ etoolbox, xparse} 
\usepackage{array}
\usepackage[normalem]{ulem}
\usepackage{amstext}

\hypersetup{colorlinks=true,linkcolor=blue,citecolor=magenta,filecolor=magenta,
  urlcolor=cyan}
\newcommand{\mg}{\textsc{MG5\_aMC}}

\newcommand{\mo}{\textsc{MicrOMEGAs}}

\newcommand{\spheno}{\textsc{SPheno}}
\newcommand{\be}{\begin{equation}}
\newcommand{\ee}{\end{equation}}
\def\bsp#1\esp{\begin{split}#1\end{split}}
\def\bpm{\begin{pmatrix}}
\def\epm{\end{pmatrix}}
\newcommand{\bea}{\begin{eqnarray}}
\newcommand{\eea}{\end{eqnarray}}

{\title{$E_6$ Motivated UMSSM Confronts Experimental Data}

\author[a]{Mariana~Frank}
\author[b,c]{\!\!, Ya\c{s}ar Hi\c{c}y\i lmaz}
\author[b]{Stefano Moretti}
\author[a]{\! and \"{O}zer \"{O}zdal}

\emailAdd{mariana.frank@concordia.ca}
\emailAdd{Y.Hicyilmaz@soton.ac.uk}
\emailAdd{S.Moretti@soton.ac.uk}
\emailAdd{ozerozdal@gmail.com}

\affiliation[a]{Department of Physics, Concordia University,
	7141 Sherbrooke St. West, Montreal, Quebec, Canada H4B 1R6}
\affiliation[b]{School of Physics $\&$ Astronomy, University of Southampton, Highfield, Southampton SO17 1BJ,UK}
\affiliation[c]{Department of Physics, Bal\i kesir University, TR10145, Bal\i kesir, Turkey}

\abstract{We  test $E_6$ realisations of a generic $U(1)'$ extended Minimal Supersymmetric Standard Model (UMSSM), parametrised in terms of  the  mixing angle pertaining to the new $U(1)'$ sector, $\theta_{E_6}$, against all currently available  data, from space to ground experiments, from low to high energies. We find that experimental constraints are very restrictive and indicate that large gauge kinetic mixing and  $\theta_{E_6}\approx -\pi/3$ are required within this theoretical construct to achieve compliance with current data. The consequences are twofold. On the one hand, large gauge kinetic mixing implies that the $Z'$  boson emerging from the breaking of the additional $U(1)'$ symmetry is rather wide since it decays mainly into $WW$ pairs. On the other hand, the preferred  $\theta_{E_6}$ value calls for a rather specific $E_6$ breaking pattern different from those commonly studied. We finally delineate potential signatures of the emerging UMSSM scenario in both Large Hadron Collider (LHC) and   in Dark Matter (DM) experiments.}

\begin{document}

\maketitle
\flushbottom

\section{Introduction}
\label{sec:intro}

After the observation of a Standard Model (SM)-like Higgs boson by ATLAS \cite{Aad:2012tfa} and CMS \cite{Chatrchyan:2012xdj} in 2012, almost all ongoing and planned observational or collider experiments have been  concentrating on searching for New Physics (NP). Undoubtedly, Supersymmetry (SUSY) is one of the most studied NP theories at these experiments, since it has remarkable advantages. In SUSY theories, the stability problem of the hierarchy between the Electro-Weak (EW) and Planck scales is solved by introducing new particles, differing by half a spin unit from the SM ones, thereby onsetting a natural cancellation between otherwise divergent boson and fermion loops in a Higgs mass or self-coupling. Furthermore,  since it relates the latter to the strength of the gauge boson couplings,  SUSY predicts a naturally light Higgs boson in its spectrum, indeed compatible with the discovered 125 GeV Higgs boson. Also,
SUSY is able to generate dynamically the Higgs potential required for  EW Symmetry Breaking (EWSB), which is instead enforced by hand in the SM. Finally,  another significant motivation for SUSY  is the natural Weakly Interacting Massive Particle (WIMP) candidate predicted in order to solve the DM puzzle, in the form of the Lightest Supersymmetric Particle (LSP). 

Though  SUSY also has the  key property of enabling gauge coupling unification, this requires rather light stops (the counterpart of the SM top quark chiral states), though,  at odds with the fact that  a 125 GeV SM--like Higgs boson  requires such stops to be rather heavy within  the Minimal Supersymmetric Standard Model (MSSM), which is the simplest SUSY extension of the SM, thereby creating an unpleasant fine tuning problem. Another phenomenological flaw of the MSSM is that, in the case of universal soft-breaking terms and the lightest neutralino as a DM candidate, the constraints from  colliders, astrophysics and rare decays have a significant impact on the parameter space of the MSSM \cite{Roszkowski:2014iqa}, such that  the MSSM, in its   constrained (or universal) version, is almost ruled out under these circumstances \cite{Abdallah:2015hza}. Moreover, the MSSM  has some theoretical drawbacks too, such as the so-called $ \mu $ problem and massless neutrinos. The aforementioned flaws of the MSSM are  motivations for non-minimal SUSY scenarios \cite{soton411745}.

Among these, UMSSMs, which have been broadly worked upon the literature, are quite popular 
\cite{Demir:2005ti,Barr:1985qs,Hewett:1988xc,Cvetic:1995rj,Cleaver:1997nj,Cleaver:1997jb,Ghilencea:2002da,King:2005jy,Diener:2009vq,Langacker:2008yv,Frank:2013yta,Frank:2012ne,Demir:2010is,Athron:2015tsa,Athron:2009bs,Athron:2009ue,Athron:2012sq,Athron:2012pw,Athron:2014pua,Athron:2015vxg,Athron:2016gor,Athron:2016qqb,Athron:2016fuq}. In the SUSY framework, these models can dynamically generate  the $ \mu $ term at the EW scale \cite{Suematsu:1994qm,Jain:1995cb,Nir:1995bu}  while even the non-SUSY versions of these are able to provide solutions for DM \cite{Okada:2010wd,Okada:2016tci,Okada:2016gsh,Agrawal:2018vin}, the muon anomaly \cite{Allanach:2015gkd} and baryon leptogenesis \cite{Chen:2011sb,Heeck:2011wj}. The right-handed neutrinos are also allowed in the superpotential to build a see-saw mechanism for neutrino masses if the extra $ U(1)$  symmetry arises from the breaking pattern of the $ E_{6} $ symmetry \cite{Keith:1996fv}. Moreover, such $ E_{6} $ motivated UMSSMs meet the anomaly cancellation conditions by heavy chiral states in the fundamental  \textbf{27} representation. 

Since there is an extra gauge boson, so-called $ Z^\prime $ boson, as well as new  SUSY particles in their  spectrum,   
UMSSM have a  richer collider phenomenology than the MSSM. Promising signals for a $ Z' $ state at the LHC would emerge from searches for heavy resonances decaying into a pair of  SM particles in Drell-Yan (DY) channels. The most stringent lower bound on the $ Z' $ mass has been set  by ATLAS in the di-lepton channel  as $ 4.5 $ TeV for an $ E_{6} $ motivated $ \psi $ model \cite{Aad:2019fac}.  Such heavy resonance searches rely upon the analysis of the  narrow Breit-Wigner (BW) line shape. In the case of the $ Z' $ boson with large decay width $ \Gamma({Z'}) $ this analysis becomes inappropriate because the signal appears as a broad shoulder spreading over the SM background instead of a narrow BW shape \cite{Accomando:2019ahs}. Furthermore, the emerging shape can be affected by a large (and often negative) interference between the broad signal and SM background. However, there are alternative experimental approaches for wide $ Z' $ resonances in the literature \cite{Accomando:2015cfa}. In these circumstances, the stringent experimental bounds on the $ Z' $ mass could be relaxed for a $ Z^\prime $ boson with a  large width $\Gamma({Z'}) $. 

This large $ Z^\prime $ width can be obtained in several Beyond the SM
(BSM) scenarios when the $Z'$ state additionally decays into exotic particles or the couplings to the fermion families are different. In  an $ E_{6} $ motivated UMSSM, through these channels, $ \Gamma({Z'}) $ could be  as large as $ 5\% $ of the $ Z' $ mass \cite{Kang:2004bz}. However, other decay channels could come into play, such as $WW$ and/or $ hZ$ (where $h$ is the SM--like Higgs boson), could have large partial widths in the presence of  gauge kinetic mixing  between two $U(1)$  gauge groups. With this in mind, we study in this work an $ E_{6} $ motivated UMSSM in a framework where such two $ U(1) $ groups kinetically mix so as to, on the one hand, enable one to find only very specific such models compatible with all current experimental data and, on the other hand, generate a wide $Z'$ which in turn allows for $Z'$ masses significantly lower than the aforementioned limits,  These could onset signals probing such constructs, at both the LHC and DM experiments.

The outline of the paper is the following. We will briefly introduce $ E_{6} $ motivated UMSSMs in Section \ref{sec:model}. After summarising our scanning procedure and enforcing experimental constraints in Section \ref{sec:scan}, we present our results over the surviving parameter space and discuss the corresponding particle mass spectrum in Section \ref{sec:spectrum}, including discussing DM implications. Finally, we summarise and conclude in Section \ref{sec:conclusion}.

\section{Model Description}
\label{sec:model}

In addition to the MSSM symmetry content, the  UMSSM includes an extra Abelian group, which we indicate as $U(1)'$. The most attractive scenario, which extends the MSSM gauge structure with an extra $U(1)'$ symmetry, can be realised by breaking  the exceptional
group $E_{6}$, an example of a possible Grand Unified Theory (GUT) 
\cite{Barr:1985qs,Hewett:1988xc,Cvetic:1995rj,Cleaver:1997nj,Cleaver:1997jb,Ghilencea:2002da,King:2005jy,Diener:2009vq,Langacker:2008yv,Langacker:1998tc,Athron:2012sq,Athron:2012pw,Athron:2014pua,Athron:2015vxg,Athron:2016gor,Athron:2016qqb,Athron:2016fuq,Hall:2010ix}, as follows:

\begin{equation}
E_{6}\rightarrow SO(10)\times U(1)_{\psi}\rightarrow SU(5)\times U(1)_{\chi}\times U(1)_{\psi}\rightarrow G_{{\rm MSSM}}\times U(1)',
\label{E6breaking}
\end{equation}
where $G_{{\rm MSSM}}=SU(3)_{c}\times SU(2)_{L}\times U(1)_{Y}$ is
the MSSM gauge group and $U(1)'$ can be expressed as a general
mixing of $U(1)_{\psi}$ and $U(1)_{\chi}$ as

\begin{equation}
U(1)'=\cos \theta_{E_{6}}U(1)_{\chi}-\sin\theta_{E_{6}}U(1)_{\psi}.
\label{Umixing}
\end{equation}

In this scenario, the cancellation of gauge anomalies is ensured by an anomaly free $E_{6}$ theory, which includes additional chiral supermultiplets. These additional chiral supermultiplets are assumed to be very heavy and embedded in the fundamental \textbf{27}-dimensional representations of $E_{6}$, which constitute the particle spectrum of this scenario alongside the  MSSM states and an additional singlet Higgs field $\hat{S}$ \cite{Langacker:1998tc}. The Vacuum Expectation Value (VEV) of $S$ is responsible for the breaking of the $U(1)'$ symmetry. Furthermore, $E_{6}$ scenarios are also encouraging candidates for extra $U(1)'$ models since they may arise from superstring theories \cite{Cvetic:2011iq}. Moreover, $E_{6}$ theories generally allow one to include see-saw mechanisms for  neutrino mass and mixing generation because of the presence of the right-handed neutrino in their \textbf{27} representations 
\cite{Hicyilmaz:2017nzo}. In this study, we assume that the right-handed neutrino does not affect the low energy implications and set its Yukawa coupling  to zero.

One can neglect the superpotential terms with the additional chiral supermultiplets as these exotic fields do not interact with the MSSM fields directly, their effects in the sparticle spectrum being quite suppressed by their masses. In this case, the UMSSM superpotential can be given as 
\begin{equation}
W = Y_{u}\hat{Q}\hat{H}_{u}\hat{U}+Y_{d}\hat{Q}\hat{H}_{d}\hat{D}+Y_{e}\hat{L}\hat{H}_{d}\hat{E}+h_{s}\hat{S}\hat{H}_{d}\hat{H}_{u},
\label{suppot1}
\end{equation}
where $\hat{Q}$ and $\hat{L}$ denote the left-handed chiral superfields for the quarks and leptons while $\hat{U}$, $\hat{D}$ and $\hat{E}$ stand for the right-handed chiral superfields of $u$-type quarks, $d$-type quarks and leptons, respectively. Here, $H_{u}$ and $H_{d}$ are the MSSM Higgs doublets and $Y_{u,d,e}$ are their Yukawa couplings to the matter fields. The corresponding Soft-SUSY Breaking (SSB) Lagrangian can be written as
\begin{equation}
\setstretch{2.0}
\begin{array}{ll}
-\mathcal{L}_{\cancel{\rm SUSY}} & = m_{\tilde{Q}}^2|\tilde{Q}|^2+m_{\tilde{U}}^2|\tilde{U}|^2+m_{\tilde{D}}^2|\tilde{D}|^2+m_{\tilde{E}}^2|\tilde{E}|^2+m_{\tilde{L}}^2|\tilde{L}|^2\\

& + m_{H_{u}}^2|H_u|^2+m_{H_{d}}^2|H_d|^2+m_S^2|S|^2 +\sum_a M_a\lambda_a\lambda_a\\

& + \left( A_SY_SSH_u\cdot H_d+A_tY_t\tilde{U}^c\tilde{Q}\cdot H_u+A_bY_b\tilde{D}^c\tilde{Q}\cdot H_d+A_{\tau}Y_b\tilde{L}^c\tilde{e}\cdot H_d+h.c. \right),
\end{array}
\label{SSB_UMSSM}
\end{equation}
where $m_{\tilde{Q}}$, $m_{\tilde{U}}$, $m_{\tilde{D}}$, $m_{\tilde{E}}$, $m_{\tilde{L}}$,$m_{H_{u}}$, $m_{H_{d}}$ and $m_{\tilde{S}}$ are the mass matrices of the scalar particles identified with the subindices, while $M_{a}\equiv M_{1},M_{2},M_{3},M_{4}$ stand for the gaugino masses. Further, $A_{S}$, $A_{t}$, $A_{b}$ and $A_{\tau}$ are the trilinear scalar interaction couplings. In Eq. (\ref{suppot1}), the MSSM bilinear mixing term $\mu H_{d}H_{u}$ is automatically forbidden by the extra $U(1)'$ symmetry and it is instead induced by the VEV of $S$ as $\mu = h_{S}v_{S}/\sqrt{2}$, where $v_{S}\equiv \langle S \rangle$. Employing Eqs.~(\ref{suppot1}) and (\ref{SSB_UMSSM}), the Higgs potential can be obtained as

\begin{equation}
V^{{\rm tree}}=V_{F}^{{\rm tree}}+V_{D}^{{\rm tree}}+V_{\cancel{\rm SUSY}}^{{\rm tree}}
\end{equation}
with

\begin{equation}
\setstretch{2.0}
\begin{array}{ll}
V_{F}^{{\rm tree}} & = | h_{s} |^{2} \left[| H_{u}H_{d}|^{2} + | S |^{2}\left( | H_{u}|^{2}+| H_{d}|^{2}  \right)   \right], \\
V_{D}^{{\rm tree}} & = \dfrac{g_{1}^{2}}{8}\left( | H_{u}|^{2}+| H_{d}|^{2}  \right)^{2}+\dfrac{g_{2}^{2}}{2}\left( |H_{u}|^{2}|H_{d}|^{2}-|H_{u}H_{d}|^{2}  \right) \\

& + \dfrac{g'^{2}}{2}\left( Q_{H_{u}}|H_{u}|^{2}+Q_{H_{d}}|H_{d}|^{2}+Q_{S}|S|^{2}  \right), \\

V_{\cancel{\rm SUSY}}^{{\rm tree}} & = m^{2}_{H_{u}}|H_{u}|^{2}+m_{H_{d}}^{2}|H_{d}|^{2}+m_{S}^{2}|S|^{2}+\left(A_{s}h_{s}SH_{u}H_{d}+h.c. \right),
\end{array}
\end{equation}
which yields the following tree-level mass for the lightest CP-even Higgs boson mass:
\begin{equation}
m_h^2=M_Z^2\cos^22\beta+\left(v_u^2+v_d^2\right)\left[\frac{h_S^2\sin^22\beta}{2}+g'^2\left(Q_{H_{u}}\cos^2\beta+Q_{H_{d}}\sin^2\beta\right)\right].
\label{h0mass}
\end{equation}
All MSSM superfields and $\hat{S}$ are charged under  the $U(1)_{\psi}$ and $U(1)_{\chi}$ symmetries and the charge configuration for any $U(1)'$ model can be
obtained from the mixing of $U(1)_{\psi}$ and $U(1)_{\chi}$, which
is quantified by the mixing angle $\theta_{E_{6}}$, through the
equation provided in the caption to Table \ref{charges}.
\begin{table}[ht!]
	\setstretch{1.5}
	\centering
	\begin{tabular}{|c|c|c|c|c|c|c|c|c|c|c|}
		\hline
		Model & $\hat{Q}$ & $\hat{U}^{c}$ & $\hat{D}^{c}$ & $\hat{L}$ & $\hat{E}^{c}$ & $\hat{H}_{d}$ & $\hat{H}_{u}$ & $\hat{S}$ \\
		\hline
		$ 2\sqrt{6}~U(1)_{\psi}$ & 1 & 1 & 1 &1 &1 & -2 & -2 & 4\\
		\hline
		$ 2\sqrt{10}~U(1)_{\chi}$ & -1 & -1 & 3 & 3 & -1 & -2 & 2 & 0 \\
		\hline
	\end{tabular}
	\caption{Charge assignments for $E_6$ fields satisfying $Q_{i}=Q_{i}^{\chi}\cos\theta_{E_{6}}-Q_{i}^{\psi}\sin\theta_{E_{6}}$.}
	\label{charges}
\end{table}

In addition to the singlet $S$ and its superpartner, the UMSSM also includes a new vector boson $Z'$ and its supersymmetric partner ${\tilde B}'$ introduced by the  $U(1)'$ symmetry. After the breaking of the $SU(2)\times U(1)_{Y}\times U(1)'$ symmetry spontaneously, $Z$ and $Z'$ mix to form physical mass eigenstates, so that  the $Z-Z'$ mass matrix is as follows
\begin{eqnarray}
\mathbf{M_{Z}^2} &=&
\left(
\begin{array}{cc}
M_{ZZ}^2	&	M_{ZZ'}^2	\\
M_{ZZ'}^2	&	M_{Z'Z'}^2	
\end{array}
\right)	= 
\left(
\begin{array}{cc}
2g_{1}^2\sum_{i} t_{3i}^2\left| \left\langle \phi_{i}\right\rangle \right|^{2}	&	2g_{1}g^{\prime}\sum_{i} t_{3i} Q_{i}\left| \left\langle \phi_{i}\right\rangle \right|^{2}	\\
2g_{1}g^{\prime}\sum_{i} t_{3i} Q_{i}\left| \left\langle \phi_{i}\right\rangle \right|^{2}	&	2g'^{2}\sum_{i} Q_{i}^2\left| \left\langle \phi_{i}\right\rangle \right|^{2}	
\end{array}
\right),
\label{ZZpmatrix}
\end{eqnarray}
where $t_{3i}$ is the weak isospin of the Higgs doublets or singlet while the $\left| \left\langle \phi_{i}\right\rangle \right|$'s stand for their VEVs. The matrix in Eq. (\ref{ZZpmatrix}) can be diagonalised by an orthogonal rotation and the mixing angle $\alpha_{ZZ'}$ can be written as 
\begin{eqnarray}
\tan 2\alpha_{ZZ'} &=& \frac{2M_{ZZ'}^2}{M_{Z'Z'}^2-M_{ZZ}^2}.
\label{az}
\end{eqnarray}
The physical mass states of $Z$ and $Z'$ are given by
\begin{eqnarray}
M^2_{Z,Z'} &=&
\frac{1}{2} 
\left[
M^2_{ZZ} + M^2_{Z'Z'} \mp 
\sqrt{ \left( M^2_{ZZ} - M^2_{Z'Z'}\right)^2 + 4 M^4_{ZZ'}}
\right] \,.
\end{eqnarray}
Besides  mass mixing, the theories with two Abelian gauge groups also allow for the existence of a gauge kinetic mixing term which is consistent with the $ U(1)_{Y} $ and $ U(1)' $ symmetries \cite{Holdom:1985ag,Babu:1997st,Rizzo:1998ut}:
\begin{eqnarray}
\mathcal{L}_{\rm kin} & \supset & -\frac{\kappa}{2} \hat{B}^{\mu\nu} \hat{Z}'_{\mu\nu} \, ,
\end{eqnarray}
where $ \hat{B}^{\mu\nu} $ and $ \hat{Z}'_{\mu\nu} $ are the field strength tensors of $ U(1)_{Y} $ and $ U(1)' $, while $ \kappa $ stands for the gauge 
 kinetic mixing parameter. The mixing factor can be generated at loop level by Renormalisation Group Equation (RGE) running while no such term appears at tree level \cite{Babu:1996vt}. In order to attach a physical meaning to the kinetic part of the Lagrangian, we need to remove the non-diagonal coupling of $ \hat{B}^{\mu\nu} $ and $ \hat{Z}'_{\mu\nu} $ by a two dimensional rotation:
\begin{eqnarray}
\left(
\begin{array}{c}
\hat{B}_{\mu} \\	\hat{Z}'_{\mu}
\end{array}
\right) 
&=&	
\left(
\begin{array}{cc}
1	&	-\frac{\kappa}{\sqrt{1-\kappa^2}}	\\
0	&	\frac{1}{\sqrt{1-\kappa^2}}
\end{array}
\right)
\left(
\begin{array}{c}
B_{\mu} \\	Z'_{\mu}
\end{array}
\right)	\, ,
\label{eq:kinZZprotation}
\end{eqnarray}
where $ \hat{B}_{\mu} $ and $ \hat{Z}'_{\mu} $ are original $ U(1)_{Y} $ and $ U(1)' $ gauge fields with off-diagonal kinetic terms while $ B_{\mu} $ and $ Z'_{\mu} $ do not posses such terms. Due to the transformation in Eq.~(\ref{eq:kinZZprotation}), a non-zero $ \kappa $ has a considerable effects on the $ Z' $ sector of the UMSSM. One of these is that the rotation matrix which diagonalises the mass matrix in Eq.~(\ref{ZZpmatrix}) is modified. Therefore, the mixing angle in Eq.~(\ref{az}) can be rewritten in terms of $ \kappa $ \cite{Babu:1997st}:
\begin{eqnarray}
\tan 2\alpha_{ZZ'} &=& \frac{-2\cos\chi (M_{ZZ'}^2+M_{ZZ}^2 \hat{s}_{W}\sin\chi)}{M_{Z'Z'}^2-M_{ZZ}^2 \cos^{2}\chi+M_{ZZ}^2 \hat{s}_{W}^{2}\sin^{2}\chi+2 M_{ZZ'}^2 \hat{s}_{W}\sin\chi} \, ,
\label{az2}
\end{eqnarray}
where $ \sin\chi=\kappa $ and $ \cos\chi=\sqrt{1-\kappa^2}$\footnote{In this notation,  generally used to express the kinetic mixing factor, $ \chi $ is called  the kinetic mixing angle.}. Note that the impact of $\kappa$ can be negligible only if $ M_{Z} \ll M_{Z'} $ and $ \kappa \ll 1 $. The $|\alpha_{ZZ'}| $ value is strongly bounded by EW Precision Tests (EWPTs) to be less than a few times $ 10^{-3}$. In models with gauge kinetic mixing (e.g., in leptophobic $Z'$ models), this limit could be relaxed but does not exceed significantly the $\mathcal{O}(10^{-3})$ ballpark \cite{Erler:2009jh}. The kinetic mixing also affects the interactions of the $ Z' $ boson with fermions. After applying the rotation in Eq.~(\ref{eq:kinZZprotation}), the Lagrangian term which shows $ Z $-fermion  and $ Z' $-fermion interaction can be written as  \cite{Rizzo:1998ut}:
\begin{eqnarray}
\mathcal{L}_{\rm int} & = & -\bar{\psi}_{i}\gamma^{\mu} \left[ g_{y} Y_{i}  B_{\mu} + (g_{p}Q_{i}+g_{yp}Y_{i})Z'_{\mu} \right]\psi_{i} \, ,
\label{Zfermionlag}
\end{eqnarray}
where $ g_{y} $, $ g_{p} $ and $ g_{yp} $ are the redefined gauge coupling matrix elements after absorbing the rotation in Eq.~(\ref{eq:kinZZprotation}) and they can be written in terms of original diagonal gauge couplings and the kinetic mixing parameter $ \kappa $:
\begin{eqnarray}
	\setstretch{2.0}
\begin{array}{lll}
g_y =&	\dfrac{g_{YY} g_{EE} - g_{YE} g_{EY}}{\sqrt{g_{EE}^2 + g_{EY}^2}}&
= g_1	,	\\
g_{yp} =&	\dfrac{g_{YY} g_{EY} + g_{YE} g_{EE}}{\sqrt{g_{EE}^2 + g_{EY}^2}}&
=	\displaystyle \frac{-\kappa g_1}{\sqrt{1-\kappa^2}}	,	\\
g_p	=&	\sqrt{g_{EE}^2 + g_{EY}^2} &
= \displaystyle	\frac{g'}{\sqrt{1-\kappa^2}},
\label{gE-define}
\end{array}
\end{eqnarray}
where $ g_{YY} $, $ g_{EE} $, $ g_{EY} $ and $ g_{YE} $ are the elements of non-diagonal gauge matrix obtained by absorbing the rotation in Eq.~(\ref{eq:kinZZprotation}) \cite{CIP10983}:
\begin{eqnarray}
G 
&=&	
\left(
\begin{array}{cc}
g_{YY}	&	g_{YE}	\\
g_{EY}	&	g_{EE}
\end{array}
\right)	\, .
\label{eq:Gmatrix}
\end{eqnarray}
{Even though the kinetic mixing term $\kappa $ does not enter the RGEs, it can be induced by the evolution of the gauge matrix terms shown in Eq. (\ref{eq:Gmatrix}), so that we have calculated  $ \kappa $ at a given scale by using the relations in Eq. (\ref{gE-define}).} It is also important to notice that parts of the mass mixing matrix in Eq. (\ref{ZZpmatrix}) change  in the case of kinetic mixing and the off-diagonal $ g_{EY} $ and $ g_{YE} $ enter in $ M_{ZZ'} $ as well.\\

As seen from Eqs.~(\ref{Zfermionlag})--(\ref{gE-define}), the kinetic mixing results in a shift in the $ U(1)' $ charges of the chiral superfields, which define the couplings of the $ Z' $ boson with fermions:
\begin{eqnarray}
Q^{eff}_{i} 
&=&	Q_{i} - \kappa \frac{g_1}{g'} Y_{i}	\,	.
\label{Qp}
\end{eqnarray}
Since the anomaly cancellation conditions for $ Q_{i} $ and $ Y_{i} $ in  $E_6$  models stabilises the theory, this new effective charge configuration is also anomaly free. Moreover, if one makes a special choice in the   $(\kappa, Q_{i}) $ space, the $ Z' $ boson can be exactly leptophobic 
\cite{Babu:1996vt,Chiang:2014yva,Araz:2017wbp}.

\newpage

Compared to the MSSM, the UMSSM has a richer gaugino sector which consists of six neutralinos. Their masses and mixing can be given in the $(\tilde{B}',\tilde{B},\tilde{W},\tilde{h}_{u},\tilde{h}_{d},\tilde{S})$
basis as follows:
\begin{equation}
\mathcal{M}_{\tilde{\chi}^{0}}=\left(\begin{array}{cccccc}
M_{1}' & 0 & 0 & g'Q_{H_{d}}v_{d} & g'Q_{H_{u}}v_{u} & g'Q_{S}v_S \\ 
0 &M_{1}& 0&-\dfrac{1}{\sqrt{2}}g_{1}v_{d} & \dfrac{1}{\sqrt{2}}g_{1}v_{u}& 0 \\ 
0 & 0&M_{2}&\dfrac{1}{\sqrt{2}}g_{2}v_{d}& -\dfrac{1}{\sqrt{2}}g_{2}v_{u}& 0 \\
g'Q_{H_{d}}v_{d} & -\dfrac{1}{\sqrt{2}}g_{1}v_{d}&\dfrac{1}{\sqrt{2}}g_{2}v_{d}&0&-\dfrac{1}{\sqrt{2}}h_{s}v_{u}& -\dfrac{1}{\sqrt{2}}h_{s}v_{u} \\ 
g'Q_{H_{u}}v_{u} &\dfrac{1}{\sqrt{2}}g_{1}v_{u}&-\dfrac{1}{\sqrt{2}}g_{2}v_{u}&-\dfrac{1}{\sqrt{2}}h_{s}v_S&0& -\dfrac{1}{\sqrt{2}}h_{s}v_{d} \\ 
g'Q_{S}v_S &  0 & 0 & -\dfrac{1}{\sqrt{2}}h_{s}v_{u} & -\dfrac{1}{\sqrt{2}}h_{s}v_{d} & 0
\end{array}
\right),
\end{equation}
where $M'_{1}$ is the SSB mass of $\tilde{B}'$ and the first row
and column encode the mixing of $\tilde{B}'$ with the other
neutralinos. Since the UMSSM does not have any new charged bosons, the chargino sector remains the same as that in the MSSM. Besides the neutralino sector, the sfermion mass sector also has extra contributions from the $D$-terms specific to the UMSSM. The diagonal terms of the sfermion mass matrix are modified by
\begin{eqnarray}
\Delta_{\tilde{f}} 
&=&	\dfrac{1}{2} g'Q_{\tilde{f}}(Q_{H_u}v_{u}^2+Q_{H_d}v_{d}^2+Q_{S}v_S^2),
\label{deltaf}
\end{eqnarray}
where $ \tilde{f} $ refers to sfermion flavours. It can be noticed that all neutralino and sfermion masses also depend on $ \kappa $ in the presence of kinetic mixing due to Eqs.~(\ref{gE-define}) and (\ref{Qp}) \cite{Belanger:2017vpq}.

\section{Scanning Procedure and Experimental Constraints}
\label{sec:scan}

In our parameter space scans, we have employed the $\spheno$ (version 4.0.0) package \cite{Porod:2003um} obtained with {\tt SARAH} (version 4.11.0) \cite{Staub:2008uz}. In this code, all gauge and Yukawa couplings in the UMSSM are evolved from the EW scale to the GUT scale that is assigned by the condition of gauge coupling unification, described as $ g_{1}=g_{2}=g' $. (Notice that $ g_{3} $ is allowed to have a small deviation from the unification condition, since it has the largest threshold corrections at the GUT scale \cite{Hisano:1992jj}.) After that, the whole mass spectrum is calculated by evaluating all SSB parameters along with gauge and Yukawa couplings back to the EW  scale. These bottom-up and top-down processes are realised by running the RGEs and the latter also requires boundary conditions given at $ M_{\rm GUT} $ scale.  In the numerical analysis of our work, we have performed random scans over the following parameter space of the UMSSM:\\
\begin{table}[h]
	\centering
	\setlength\tabcolsep{8pt}
	\renewcommand{\arraystretch}{1.4}
	\begin{tabular}{c|c||c|c}
		Parameter  & Scanned range & Parameter      & Scanned range \\
		\hline
		$m_0$ & $[0., 3.]$ TeV     & $h_s$    & $[0., 0.7]$\\
		$M_{1,4}/M_3$ & $[-15., 15.]$  & $v_S$  & $[1., 15.]$ TeV\\
		$M_3$        & $[0.,3.]$ TeV & $A_s$ & $[-5., 5.]$ TeV\\
		$M_2/M_3$ & $[-5.,5.]$  & $\theta_{E_6}$& $[-\pi/2, \pi/2]$ \\
		$\tan\beta$ & $[1., 50.]$   & $\kappa$ & $[-0.5,0.5]$\\
		$A_0$    & $[-5., -5.]$ TeV  &  \\
	\end{tabular}
	\caption{Scanned parameter space.}
	\label{paramSP}
\end{table}
\\
where $ m_{0} $ is the universal SSB mass term for the matter scalars while  $ M_{1}, M_{2}, M_{3}, M_{4} $ are the {non-universal} SSB mass terms of the gauginos {at the GUT scale} associated with the $ U(1)_Y $, $ SU(2)_L $, $ SU(3)_c $ and $ U(1)' $ symmetry groups, respectively. Besides, $ A_0 $ is the SSB trilinear coupling  and $ \tan\beta $ is the ratio of the VEVs of the MSSM Higgs doublets. $ A_s $ is the SSB interaction between the $ S $, $ H_u $ and $ H_d $ fields. In addition, as mentioned previously, $ \theta_{E_{6}} $ and $ \kappa $ are the $ Z-Z' $ mass mixing angle and gauge kinetic mixing parameter. Finally, we also vary the Yukawa coupling $h_s$ and $v_S$ (the VEV of $S$), which is responsible for the breaking of the $U(1)'$ symmetry.

{An $ E_6 $ based UMSSM with \textbf{27} representations can achieve unification of the Yukawa  as well as  gauge couplings at the GUT scale if $ E_6 $  is broken down to the MSSM gauge group via $SO(10)$ \cite{Gogoladze:2011ce}. (The non-universality of the gaugino masses can also be tolerated  when $SO(10)$ is broken down to a Pati-Salam gauge group \cite{Gogoladze:2009ug,Altin:2017sxx}.) However, starting from the Yukawa couplings, one needs to fit the top, bottom and tau masses in presence of very stringent experimental constraints. Despite the fact that the general UMSSM framework can be consistent with the latter (as well as with the discovered Higgs boson mass) \cite{Hicyilmaz:2016kty}, the ensuing requirements on the parameter space are extremely restrictive, so that,   for our analysis,  we do not assume any  $ t-b-\tau $ (or  even $ b-\tau $) Yukawa coupling unification.}

In order to scan the parameter space efficiently, we use the Metropolis-Hasting algorithm \cite{Belanger:2009ti}. After data collection, we implement Higgs boson and sparticle mass bounds \cite{Chatrchyan:2012xdj,Tanabashi:2018oca} as well as constraints from Branching Ratios (BRs) of $B$-decays such as $ {\rm BR}(B \rightarrow X_{s} \gamma) $ \cite{Amhis:2012bh}, $ {\rm BR}(B_s \rightarrow \mu^+ \mu^-) $ \cite{Aaij:2012nna} and $ {\rm BR}(B_u\rightarrow\tau \nu_{\tau}) $ \cite{Asner:2010qj}. We also require that the predicted relic density of the neutralino LSP agrees within 20\% (to conservatively allow for uncertainties on the  predictions)  with  the  recent Wilkinson Microwave Anisotropy Probe (WMAP) \cite{Hinshaw:2012aka} and Planck  results, $\Omega_{\rm CDM} h^2 =  0.12$ \cite{Ade:2013zuv,Aghanim:2018eyx}.  The relic density of the LSP and scattering cross sections for direct detection experiments are calculated with $\mo$ (version 5.0.9) \cite{Belanger:2018mqt}. The experimental constraints can be summarised as follows:
\begin{equation}
\setstretch{1.8}
\begin{array}{l}
m_h  = 123-127~{\rm GeV} ({\rm{and~SM-like~couplings}}),
\\
m_{\tilde{g}} \geq 1.8~{\rm TeV},
\\
0.8\times 10^{-9} \leq{\rm BR}(B_s \rightarrow \mu^+ \mu^-)
\leq 6.2 \times10^{-9} \;(2\sigma~{\rm tolerance}),
\\
m_{\tilde{\chi}_{1}^{0}} \geq 103.5~{\rm GeV}, \\
m_{\tilde{\tau}} \geq 105~{\rm GeV}, \\
2.99 \times 10^{-4} \leq
{\rm BR}(B \rightarrow X_{s} \gamma)
\leq 3.87 \times 10^{-4} \; (2\sigma~{\rm tolerance}),
\\
0.15 \leq \dfrac{
	{\rm BR}(B_u\rightarrow\tau \nu_{\tau})_{\rm UMSSM}}
{{\rm BR}(B_u\rightarrow \tau \nu_{\tau})_{\rm SM}}
\leq 2.41 \; (3\sigma~{\rm tolerance}), \\
0.0913 \leq \Omega_{{\rm CDM}}h^{2} \leq 0.1363~(5\sigma~{\rm tolerance}).
\label{constraints}
\end{array}
\end{equation}

As discussed in the previous section, the kinetic mixing affects the $ Z-Z' $ mixing matrix and adds new terms related to the off-diagonal gauge matrix elements $ g_{EY} $ and $ g_{YE} $ into the mixing term $ M_{ZZ'}$. Furthermore, the mixing angle could be enhanced near or beyond the EWPT bounds. The main reason is that the new $ M_{ZZ'}$ element includes the term with proportional to $ g_{EY} Q_{S}^2 v_{S}^2 $. Therefore, one must take a specific $ g_{EY} $ range if one wants to avoid violating the EWPT limits for $ \alpha_{ZZ'} $. In our analysis, we allow this range as $ g_{EY} \sim \mathcal{O}(10^{-3}) $  to obtain a large (but compatible with EWPTs) $ \alpha_{ZZ'} $, as $ \Gamma({Z' \to WW}) $ and $\Gamma({Z'\to Zh})$ are very sensitive to this coupling. 
In order to account for EWPTs, we have parameterised the latter through the  EW oblique parameters $S, T$ and $ U$ that are obtained from the $\spheno$ output \cite{Altarelli:1990zd,Peskin:1990zt,Peskin:1991sw,Maksymyk:1993zm,Baak:2014ora}.

In the case that $\Gamma(Z^{\prime})/M_{Z^{\prime}}$ is large\footnote{Notice that we have put a bound on the total width of the $ Z' $ boson, $\Gamma(Z^{\prime})\lesssim M_{Z^{\prime}}/2$, so as to avoid unphysical resonance behaviours \cite{Grassi:2001bz}.}, the LHC limits on the $Z'$ boson mass and couplings, which are produced under the assumption of Narrow Width Approximation (NWA), cannot be applied, as interference effects are not negligible \cite{Accomando:2013sfa,ACCOMANDO:2013ita}. Therefore, here, we define the $Z'$ Signal (S) as the difference between $\sigma(pp \to \gamma,Z,Z' \to ll)$ and the SM Background (B) $\sigma(pp \to \gamma,Z \to ll)$, where $l=e,\mu$. The corresponding cross section values have been calculated by using $\mg$ (version 2.6.6) \cite{Alwall:2014hca} along with the leading-order set of NNPDF 2.3 parton densities \cite{Ball:2012cx}.

The following  list  summarises the relation between colours and constraints imposed in our forthcoming plots.

\begin{itemize}
	\item Grey: Radiative EWSB (REWSB) and  neutralino LSP.
	\item Red: The subset of grey  plus  Higgs boson mass and coupling constraints, SUSY particle mass bounds and EWPT requirements.
	\item Green:  The subset of red  plus  $B$-physics constraints.
	\item Blue: The subset of green  plus WMAP constraints on the relic abundance of the neutralino LSP (within $5\sigma$).
	\item Black: The subset of blue  plus  exclusion limits at the LHC from 
	$Z'$  direct searches via $ pp\to Z' \to ll  $ and $ pp\to Z' \to WW$.
\end{itemize}

We further discuss the application of these limits in the next section. We ignore here  $(g-2)_\mu$ constraints, as we can anticipate that the corresponding predictions in our $E_6$ inspired UMSSM are consistent with the SM, due to the fact that the relevant slepton and sneutrino masses are rather heavy and so is  the $Z'$ mass.

\section{Mass Spectrum and Dark matter}
\label{sec:spectrum}
\begin{figure}[H]
	\centering
	\subfigure{\includegraphics[scale=0.45]{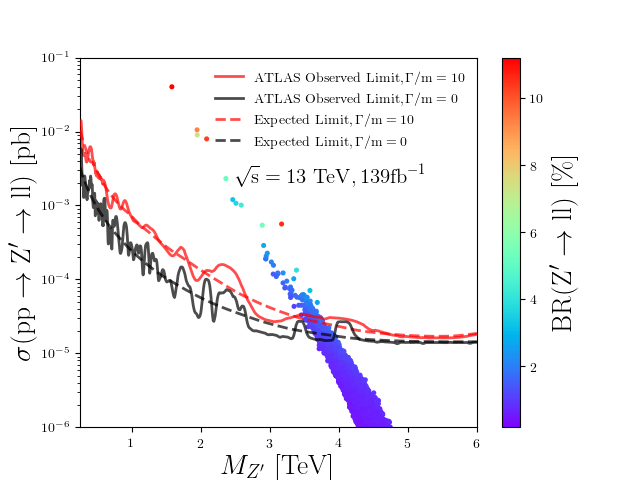}}
	\subfigure{\includegraphics[scale=0.45]{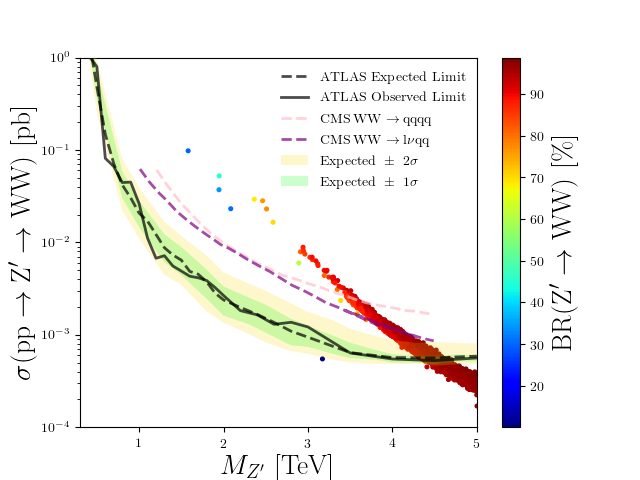}}
	\caption{The $ Z' $ boson mass limits on $\sigma(pp \to Z' \to ll)$ vs $M_{Z'}$ (left panel) and  $\sigma(pp \to Z' \to WW)$ vs $M_{Z'}$ (right panel). The experimental exclusion curves obtained by the ATLAS \cite{Aad:2019fac,Aaboud:2017fgj} and CMS \cite{Sirunyan:2017acf,Sirunyan:2018iff} collaborations are showed against the results of our scan colour coded in terms of the relevant $Z'$ BR. }
	\label{fig:sigma_mZp}
\end{figure}

This section will start by presenting our results for the $ Z' $ mass and coupling bounds (in a  large $ \Gamma({Z'}) $ scenario) and how these can be related to the fundamental charges of an $E_6$ inspired UMSSM, then, upon introducing the LHC constraints affecting the SUSY sector, it will move on to discuss the DM phenomenology in astrophysical conditions.

Fig. \ref{fig:sigma_mZp} shows the comparison of the experimental limits on the $ Z' $ boson mass and cross section (hence some coupling combinations) as obtained from direct searches in the processes  $ pp \to ll $ at $ \mathcal{L}=137~{\rm{fb}}^{-1} $ \cite{Aad:2019fac}  and $ pp \to WW $  at $ \mathcal{L}=36~{\rm{fb}}^{-1} $ \cite{Aaboud:2017fgj,Sirunyan:2017acf,Sirunyan:2018iff}. All points plotted here satisfy all constraints that are coded  ``Blue" in the previous section while the actual colours display the BR of the related  $ Z' $ boson decay channel. According to our results, in the left panel, we find that the $ Z' $ boson mass cannot be smaller than   $ 3.5 $ TeV in the light of the ATLAS dilepton results \cite{Aad:2019fac}. Indeed, it is thanks to the gauge kinetic mixing effects on the $ U(1)' $ charges and the negative interference onset by the wide $ Z' $ with the SM background that we are able to obtain this lower limit, as the ATLAS results  \cite{Aad:2019fac} reported a lower limit at $ 4.5 $ TeV (e.g., for an  $E_6$  based $ \psi $ model). Furthermore, as can be seen from the right panel, the ATLAS results on the $Z' \to WW$ channel \cite{Aaboud:2017fgj}, when taken within $ 2\sigma $,  put a lower $Z'$ mass limit at $ M_{Z'} \gtrsim 4  $ TeV. This lower bound is somewhat relaxed by  some CMS results also shown in the same plot, down to 3.5 TeV. In the reminder of this work, therefore, we use the $ Z' $ boson mass allowed by all $Z'$ direct searches in the dilepton and diboson channels as being $ M_{Z'} \gtrsim 4  $ TeV. 

\begin{figure}[H]
	\centering
	\subfigure{\includegraphics[scale=0.36]{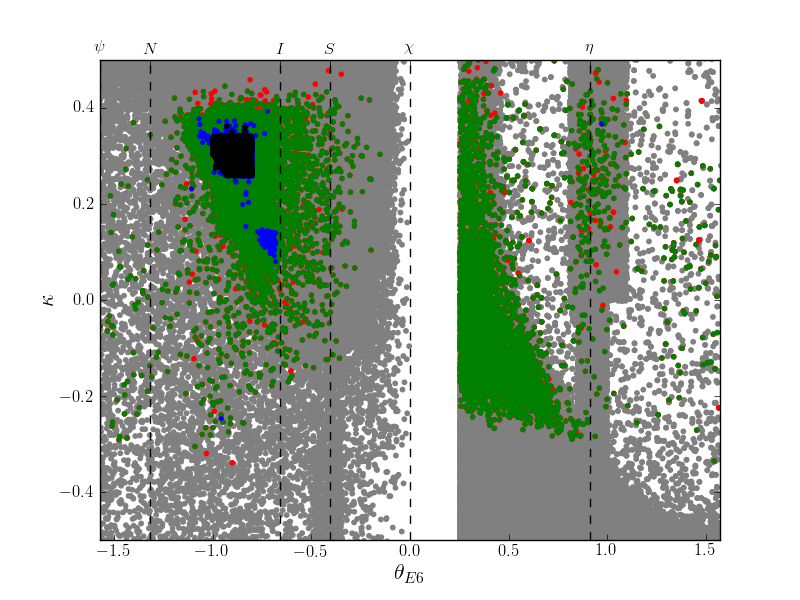}}
	\subfigure{\includegraphics[scale=0.45]{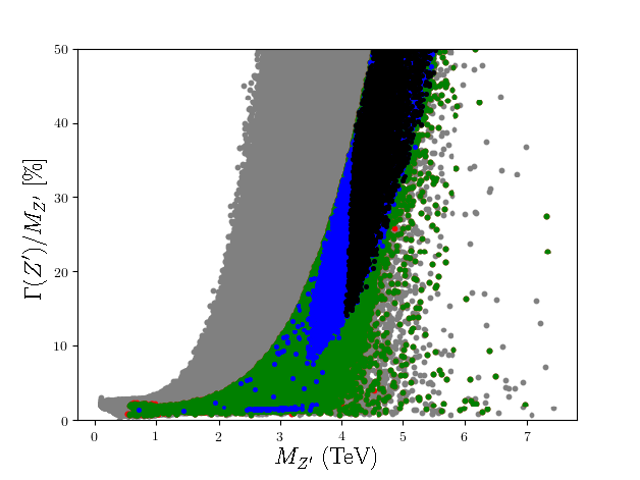}}
	\caption{The gauge kinetic mixing parameter $\kappa$ versus $U(1)'$ charge mixing angle $\theta_{E_6}$ 
		(left panel) and the $Z'$ width-to-mass ratio $ \Gamma(Z^{\prime})/M_{Z^{\prime}} $ vs the $Z'$ mass $M_{Z'}$ (right panel).  Our colour convention is as listed at the end of Section \ref{sec:scan}. The vertical dashed lines in the left panel corresponds to well-known $E_6$  realisation with defined $\theta_{E_6}$ choices.}
	\label{fig:E6_k}
\end{figure}

In Fig. \ref{fig:E6_k} we present our results in plots showing the gauge kinetic mixing parameter versus the $U(1)'$ charge mixing angle, i.e., on the plane ($\theta_{E_{6}}, \kappa)$ (left panel), and  the $ Z' $ boson mass  versus the ratio of its total decay width over the former, i.e., on the plane $  (M_{Z^{\prime}}, \Gamma(Z^{\prime})/M_{Z^{\prime}}) $ (right panel). The former plot shows that the parameter space of the $\theta_{E_{6}}$ mixing angle, which also defines the effective charge of $ U(1)' $, is constrained severely when we apply all limits mentioned in Section \ref{sec:scan}. We see  that  $\theta_{E_{6}}$ values are found in the interval $ \left[ -1, -0.8\right]  $ radians while the corresponding $ \kappa $ values are found in $ \left[ 0.2, 0.4\right]$. We notice that such solutions do not accumulate against any of the most studied $E_6$ realisations, known as   $\psi, N, I, S, \chi$ and $\eta$
\cite{Tanabashi:2018oca}. The latter plot indeed makes the point that wide $Z'$ states are required  to evade LHC limits from $Z'$ direct searches, with values of the width being no less than 15\% or so of the mass. The right panel shows that $ \Gamma(Z^{\prime})/M_{Z^{\prime}} $ can drastically increase with large $ M_{Z^{\prime}} $. This is due to the fact that the decay width $ \Gamma(Z' \to WW) $ is proportional to $ ( M_{Z'}^5/M_{W}^4) $ as well as $ \sin^{2}\alpha_{ZZ'} $ \cite{Bandyopadhyay:2018cwu}. (Recall that the ``Black" points here include the constraints drawn from the previous figure.)

\begin{figure}[H]
	\centering
	\subfigure{\includegraphics[scale=0.36]{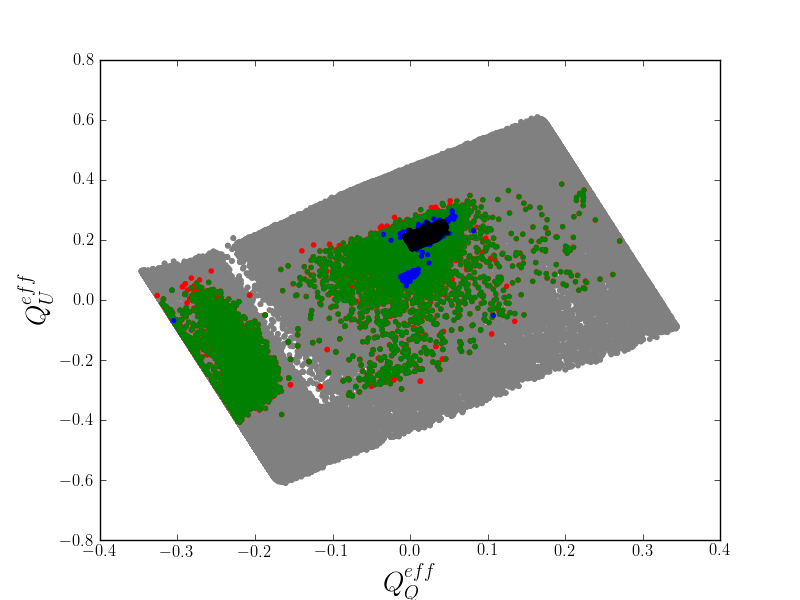}}
	\subfigure{\includegraphics[scale=0.36]{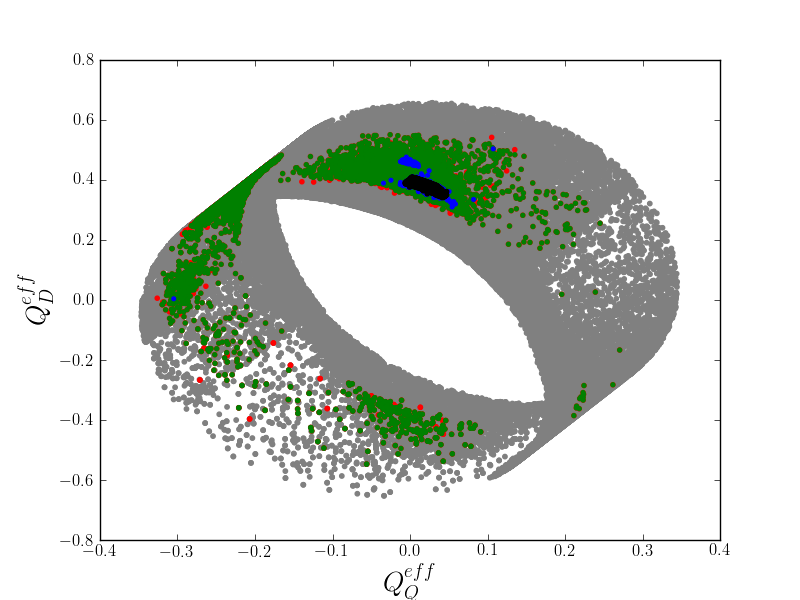}}
	\subfigure{\includegraphics[scale=0.36]{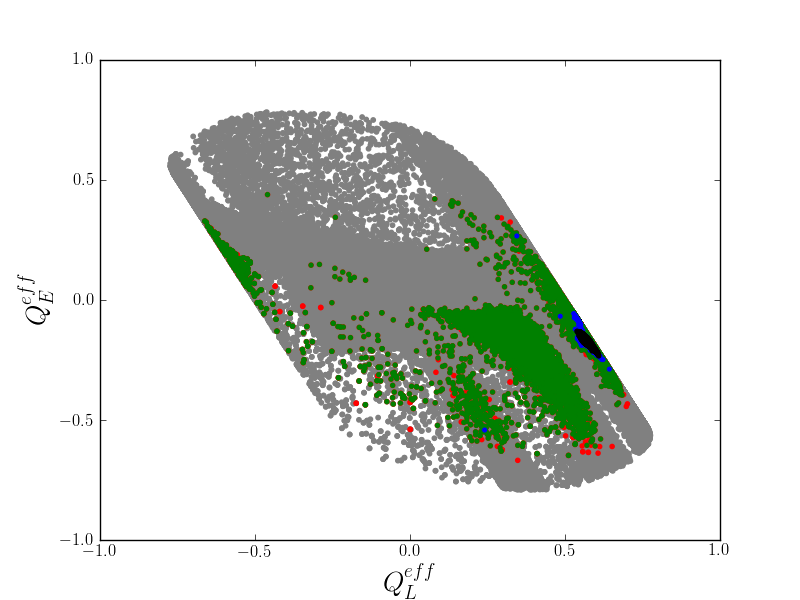}}
	\subfigure{\includegraphics[scale=0.45]{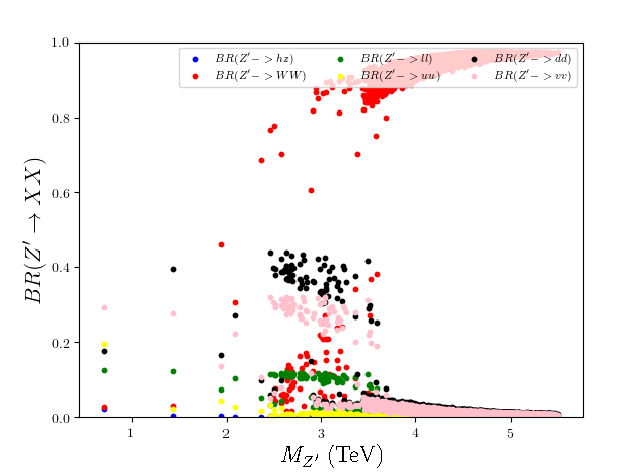}}
	\caption{The distributions of the effective $U(1)'$ charges for quarks and leptons over the following planes: 
		($Q^{eff}_{Q},Q^{eff}_{U}$) (top left),  ($Q^{eff}_{Q},Q^{eff}_{D}$) (top right) and  ($Q^{eff}_{L},Q^{eff}_{E}$) (bottom left). In the bottom right plot we show the BRs of the $ Z' $ for different decay channels, BR$(Z' \to XX) $ as a function on $M_{Z'}$, where $XX$ represents a SM two-body final state. Our colour convention is as listed at the end of Section \ref{sec:scan} and the bottom right panel contains only the ``Blue" points in the other panels.}
	\label{fig:q_q}
\end{figure}

The solutions in the $(\theta_{E_{6}},\kappa)$ region which we have just seen  have special $ U(1)'$ effective charge configurations, are presented in Fig.~\ref{fig:q_q}. Herein,  we show such charges, as given in Eq.~(\ref{Qp}), for left and right chiral fermions by visualising our scan points over the planes $(Q^{eff}_{Q},Q^{eff}_{U}$),  $(Q^{eff}_{Q},Q^{eff}_{D})$ and $(Q^{eff}_{L},Q^{eff}_{E})$. As seen from the top left and right panels,  when we take all experimental constrains into consideration (``Black" points), the family universal effective $ U(1)' $ charges for left handed ($ Q^{eff}_{Q} $) quarks are always very small, with the right handed up-type ($ Q^{eff}_{U}$) quark charges smaller than those of the  right handed down-type ($ Q^{eff}_{D}$) ones. As for leptons, it is the left handed ($ Q^{eff}_{L} $) charges which are generally larger than the right handed ones $( Q^{eff}_{E}) $ (as shown in the bottom left panel of the figure).  This pattern builds up the
distribution of fermionic BRs seen in the bottom right panel of the figure, as the  partial decay width of the $ Z' $ into fermions  $f$, $ \Gamma(Z' \to ff), $ is proportional to $ M_{Z'}  ({Q^{eff}_{\rm left}}^2 + {Q^{eff}_{\rm right}}^2) $ \cite{Kang:2004bz}. However, such a  BR$(Z' \to XX) $ distribution is actually dominated by $Z'\to WW$ decays over most of the $M_{Z'}$ range (with the companion $Z'\to Zh$ channel always subleading), 
given that, for  large $ Z' $ masses, as mentioned, $ \Gamma(Z' \to WW) $  is proportional to $ M_{Z'}^5/M_{W}^4 $, hence the rapid rise up to $98\%$ with increasing $M_{Z'}$, particularly so from 4 TeV onwards (notice that these decay distributions have been produced by the ``Blue" points appearing in the other panels). It is thus not surprising that the most constraining search for the $Z'$ of $E_6$ inspired UMSSM scenarios is the diboson one, rather than the  dilepton one (limitedly to the case of its SM decay channels).

\begin{figure}[H]
	\centering
	\subfigure{\includegraphics[scale=0.46]{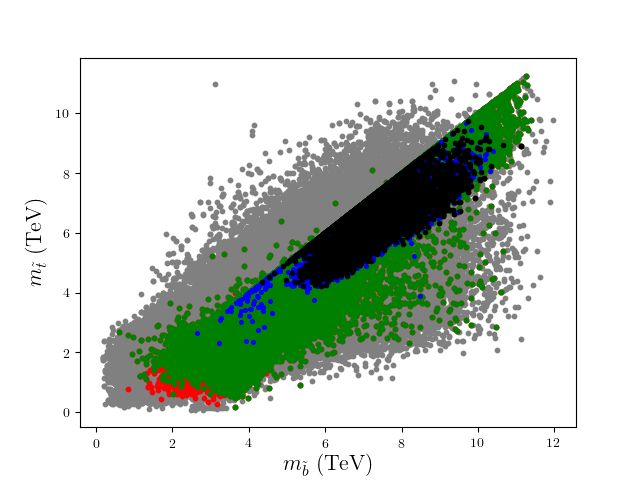}}
	\subfigure{\includegraphics[scale=0.46]{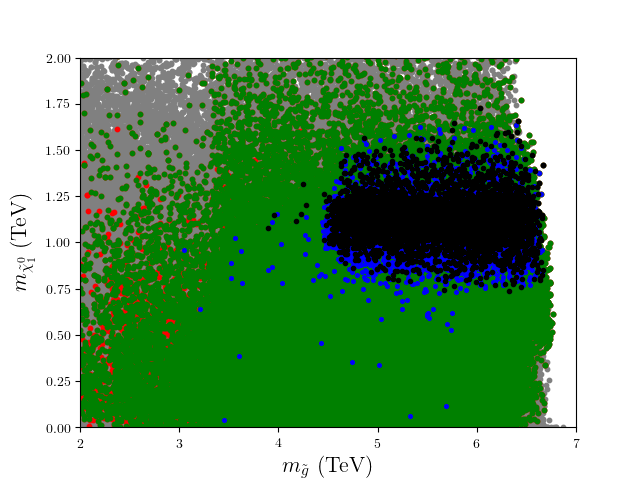}}
	\subfigure{\includegraphics[scale=0.36]{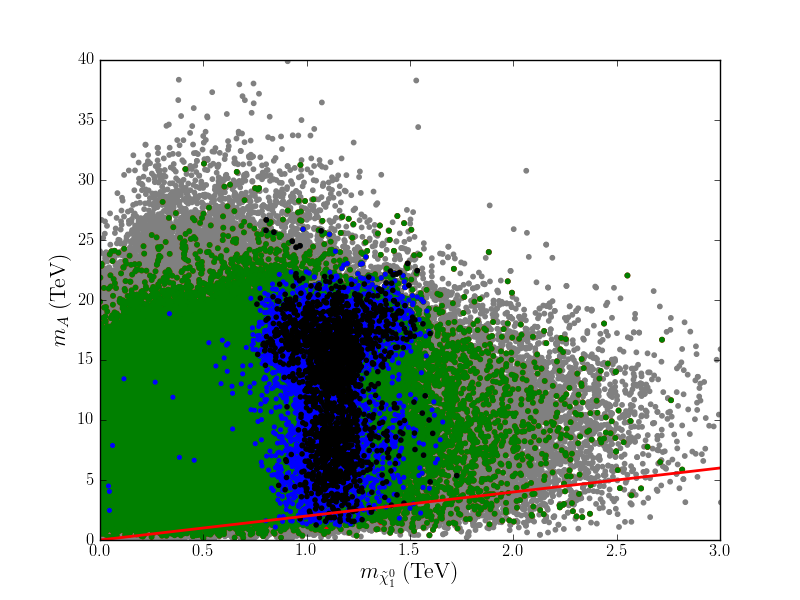}}
	\subfigure{\includegraphics[scale=0.46]{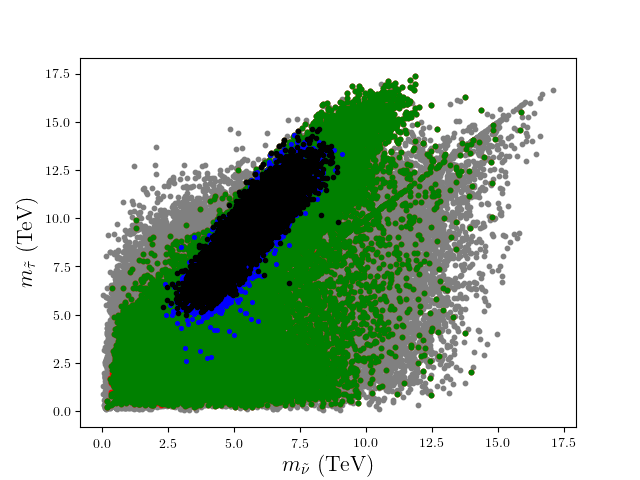}}
	\caption{The mass spectrum of Higgs and SUSY states over the following planes:  $(m_{\tilde{b}},m_{\tilde{t}})$ (top left), 
		$(m_{\tilde{g}},m_{\tilde{\chi}_{1}^{0}})$ (top right), 
		$(m_{\tilde{\chi}_{1}^{0}},m_A)$ (bottom left) and $(m_{\tilde{\nu}},m_{\tilde{\tau}})$ (bottom right). Our colour convention is as listed at the end of Section \ref{sec:scan}.}
	\label{fig:mb_mt}
\end{figure}

We now move on to study the other two sectors of our $U(1)'$ construct, namely, the spectrum of Higgs and SUSY particle masses.
A selection of these is presented in Fig.~\ref{fig:mb_mt} with plots over the following mass combinations (clockwise):  ($m_{\tilde{b}},m_{\tilde{t}}$),  ($m_{\tilde{g}},m_{\tilde{\chi}_{1}^{0}}$), ($m_{\tilde{\chi}_{1}^{0}},m_A$) and ($m_{\tilde{\nu}},m_{\tilde{\tau}}$). The colour coding is the same as 
the one  listed at the end of Section \ref{sec:scan}. As seen from the top left and right panels of the figure, the  SUSY mass spectrum of the allowed parameter region (i.e., the ``Black" points)  is quite heavy with the lower limit on stop, sbottom and gluino masses of about $ 4 $ TeV. The reason for the  large sfermions mass arises from the fact that the contributions of the $ U(1)' $ sector to such  masses are proportional to $ v_{S}^2 $, which also determines the mass of the $ Z'$. Therefore, the experimental limits on the $ Z' $ mass in Fig.~\ref{fig:sigma_mZp} in turn drive those on the sfermion masses. The bottom left panel shows that  the LSP (neutralino) mass should be $ 0.8$ TeV $\lesssim m_{\tilde{\chi}_{1}^{0}} \lesssim 1.7 $ TeV (the extremes of the ``Black" point distribution).  
In this plot, the solid red  line shows the points with  $m_A=2m_{\tilde{\chi}_{1}^{0}}$, condition onsetting the dominant resonant DM annihilation via $A$ mediation, 
so that very few  solutions (to WMAP data) are found below it. As for the stau masses, see bottom right frame, 
these are larger than the sneutrino ones (again, see the ``Black" points), both well in the TeV range. In summary, both the Higgs and SUSY (beyond the LSP)  mass spectrum is rather heavy, thus explaining the notable absence of non-SM decay channels for the $Z'$, as already seen.

\begin{figure}[H]
	\centering
	\subfigure{\includegraphics[scale=0.46]{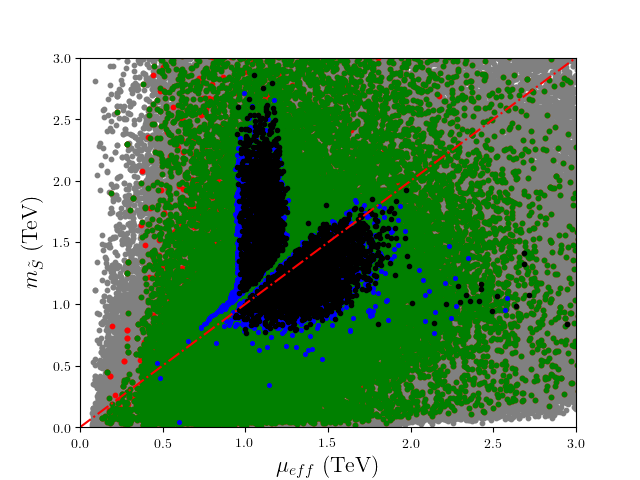}}
	\subfigure{\includegraphics[scale=0.46]{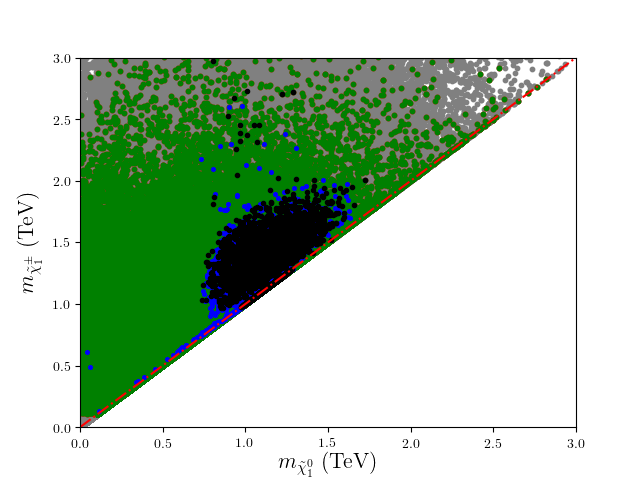}}
	\subfigure{\includegraphics[scale=0.46]{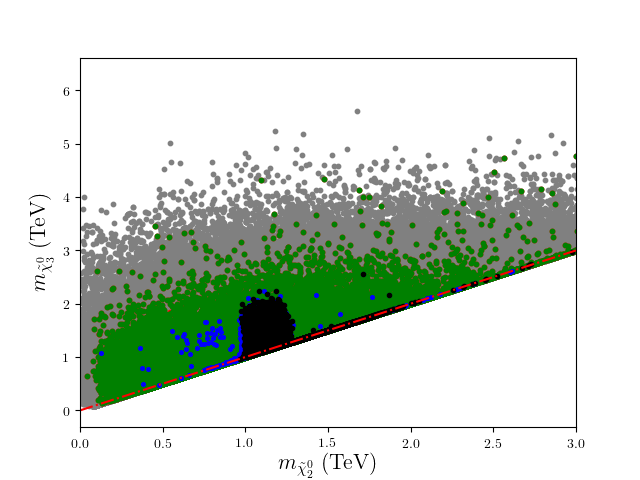}}
	\subfigure{\includegraphics[scale=0.46]{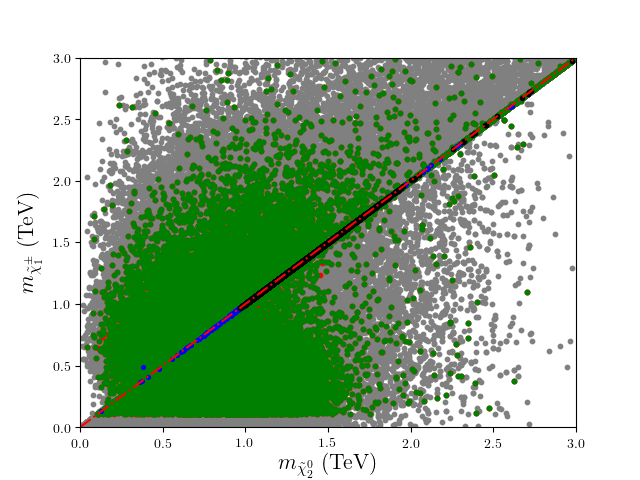}}
	\caption{The mass spectrum of chargino and neutralino states over the following planes: ($\mu_{eff}, m_{\tilde{S}}$) (top left),     ($m_{\tilde{\chi}_{1}^{0}},m_{\tilde{\chi}_{1}^{\pm}}$) (top right),
		($m_{\tilde{\chi}_{2}^{0}},m_{\tilde{\chi}_{0}^{3}}$) (bottom left) 
		and ($m_{\tilde{\chi}_{2}^{0}},m_{\tilde{\chi}_{1}^{\pm}}$) (bottom right). Our colour convention is as listed at the end of Section \ref{sec:scan}.}
	\label{fig:chi_mu}
\end{figure}

In Fig.~\ref{fig:chi_mu} we illustrate the neutralino and chargino mass spectrum, also in relation to the effective $\mu$ parameter, $\mu_{eff}$, using  plots over the following parameter combinations ($\tilde S$ being the singlino): ($\mu_{eff},m_{\tilde{S}})$, $(m_{\tilde{\chi}_{1}^{0}}, m_{\tilde{\chi}_{1}^{\pm}})$, $(m_{\tilde{\chi}_{2}^{0}},m_{\tilde{\chi}_{0}^{3}})$ and $(m_{\tilde{\chi}_{2}^{0}},m_{\tilde{\chi}_{1}^{\pm}})$. (The colour coding is the same as in Fig.~\ref{fig:E6_k}.) Herein, (the diagonal) dot-dashed red lines indicate regions in which the displayed parameters are degenerate in value. The top left panel shows that the LSP, the neutralino DM candidate, is higgsino-like or singlino-like since the other gauginos that contribute to the neutralino mass matrix are heavier and decouple (see below). The higgsino-like DM mass can be $ 1$ TeV $\lesssim m_{\tilde{\chi}_{1}^{0}} \lesssim 1.2 $ TeV while the singlino-like DM mass can cover a wider range,  $ 0.8$ TeV $\lesssim m_{\tilde{\chi}_{1}^{0}} \lesssim 1.7 $ TeV. Further, as can be seen from the top right panel, the lightest  chargino and LSP  are largely degenerate in mass (typically, within a few hundred GeV) in the region of the higgsino-like DM mass and the chargino mass can  reach $ 3 $ TeV.  These solutions favour the chargino-neutralino coannihilation channels which reduce the relic abundance of the LSP,  such that the latter can be consistent with the WMAP bounds. (This region also yields the $A$ resonant solutions, $m_A=2m_{\tilde{\chi}_1^{0}}$, as seen from the bottom left panel of  Fig.~\ref{fig:mb_mt}.) 
The bottom left panel  illustrates the point that, for higgsino-like DM, the mass gap between the second and third lightest neutralino can be of order TeV,
though there is also a region with significant mass degeneracy.
Then, as seen from the bottom right panel, the lightest chargino and second lightest neutralino  are extremely degenerate in mass for all allowed solutions
(``Black" points). Altogether, this means that  EW associated production of mass degenerate charginos $ \tilde{\chi}_{1}^{\pm} $ and neutralinos $ \tilde{\chi}_{2}^{0} $ where $ \tilde{\chi}_{1}^{\pm} \to W\tilde{\chi}_{1}^{0} $ and $ \tilde{\chi}_{2}^{0} \to h \tilde{\chi}_{1}^{0} $ is possible for both type of higgsino- and singlino-like LSP. However, it must be said that EW production of mass degenerate neutralinos cannot be possible because of the heavy sleptons shown in the bottom right panel of Fig.~\ref{fig:mb_mt}. Hence, a potentially interesting  new production and decay mode emerges in the -ino sector, $ pp \to \tilde{\chi}_{2}^{0} \tilde{\chi}_{3}^{0} \to (h/Z)(h/Z) \tilde{\chi}_{1}^{0} \tilde{\chi}_{1}^{0} $, which could be probed at the High Luminosity LHC (HL-LHC).

\begin{figure}[H]
	\begin{center}
		\subfigure{\includegraphics[scale=0.46]{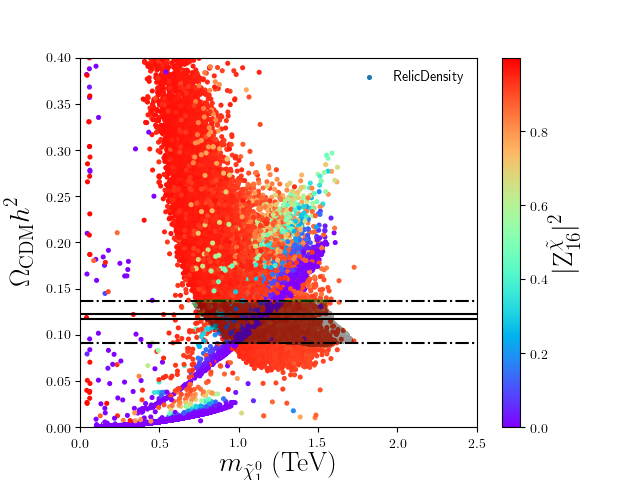}}
		\subfigure{\includegraphics[scale=0.46]{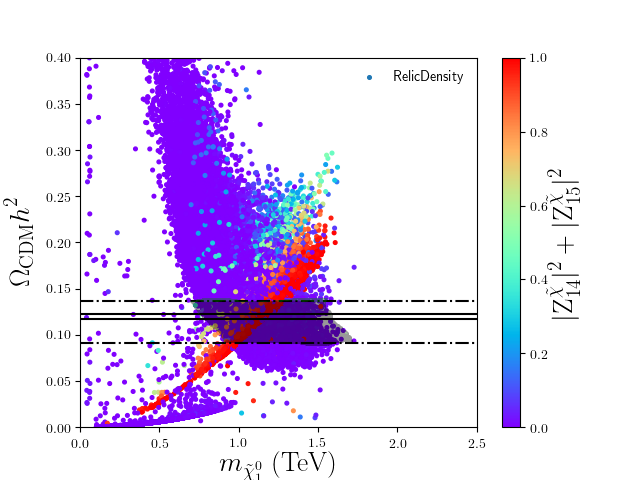}}		
		\caption{Relic density predictions for  singlino-like  (left) and higgsino-like (right) DM as a function of  the  mass  of the neutralino LSP. The colour bars show the composition of the LSP. The meaning of the horizontal lines is explained in the text.}
		\label{fig:relic_chi1}
	\end{center}
\end{figure}

Before closing, we investigate how cosmological bounds from relic density and from  DM experiments impact our solutions.
Fig. \ref{fig:relic_chi1} shows that our relic density predictions for singlino LSP (left panel) and higgsino LSP (right panel) as the DM candidate. 
The color bars show the singlino (left panel) and higgsino (right panel) compositions of LSP. (Notice that the population of points used in this plot correspond to the ``Green'' points   listed at the end of Section \ref{sec:scan}, i.e., meaning that all experimental constraints, except for DM itself and the  $Z'$ mass and coupling limits, are applied.) The dark shaded areas between the horizontal lines show where the ``Black"  points are in this figure. The dot-dashed(solid) lines indicate the WMAP bounds on the relic density of the DM candidate within a $ 5\sigma(1\sigma) $ uncertainty. The region within  the dot-dashed lines covers also   the recent Planck bounds \cite{Aghanim:2018eyx}. Altogether, the figure points to a singlino-like DM being generally more consistent with all relic density data available, though the higgsino-like one is also viable, albeit in a narrower region of parameter space, with the two solutions overlapping each other.

\begin{figure}[H]
	\centering
	\subfigure{\includegraphics[scale=0.36]{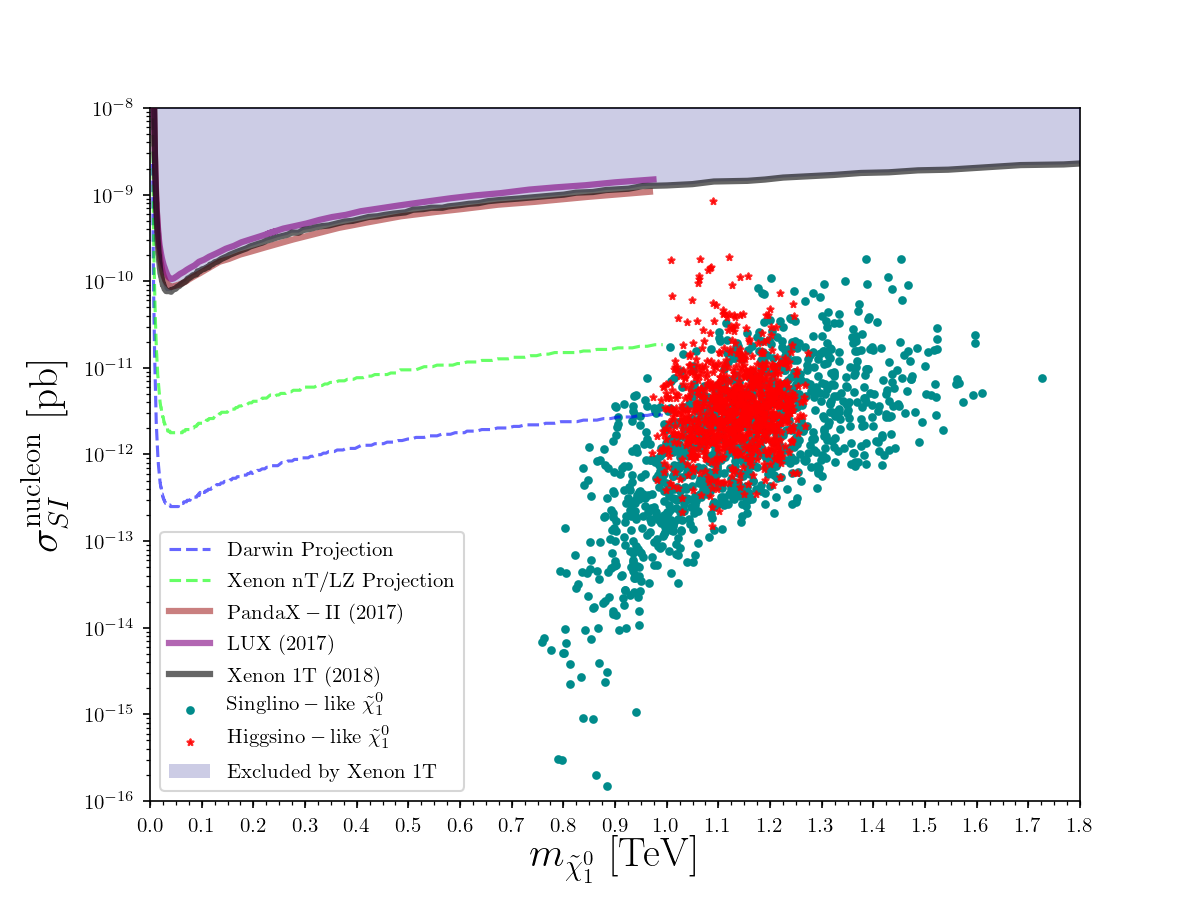}}
	\subfigure{\includegraphics[scale=0.36]{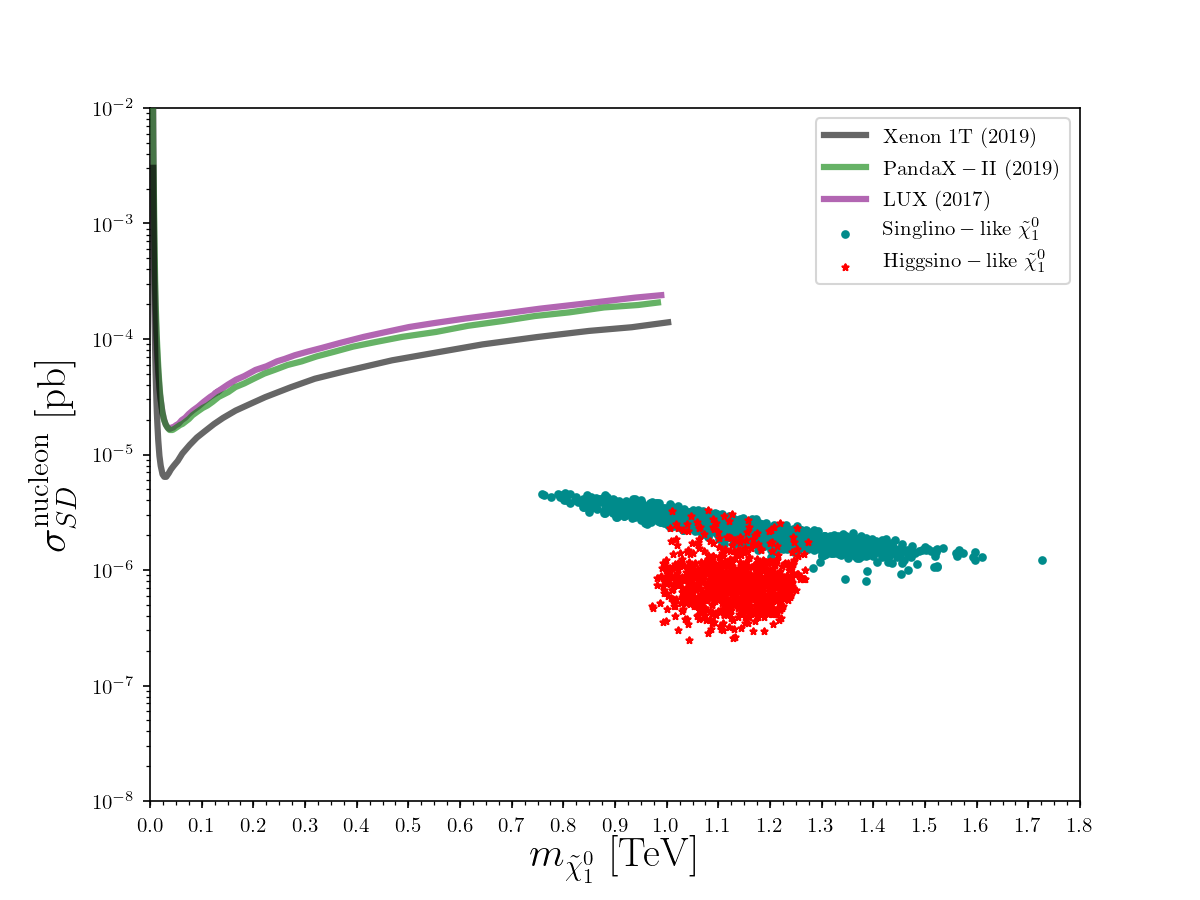}}
	\caption{DM-neutron SI (left) and  SD (right) scattering cross section  as a function of  the  mass  of the WIMP candidate 
		(neutralino LSP).  The colour bars show the composition of the LSP. Limits from current (solid) and future (dashed) experiments are also shown.}
	\label{fig:sigma_mZp2}
\end{figure}

In Fig. \ref{fig:sigma_mZp2} we depict the DM-neutron Spin-Independent (SI, left panel) and  Spin-Dependent (SD, right panel) scattering cross sections  as  functions of  the WIMP candidate mass, i.e., that of the neutralino LSP. The color codes are indicated in the legend of the panels. Here, all points  satisfy all the experimental constraints used in this work, i.e., they correspond to the ``Black'' points as described at the end of  Section \ref{sec:scan}. We represent solutions with $|Z_{16}^{\tilde\chi}|^2 > 0.6$ as singlino-like $\tilde\chi_1^0$ and show them in dark cyan colour. Likewise, solutions with $|Z_{14}^{\tilde\chi}|^2 + |Z_{15}^{\tilde\chi}|^2  > 0.6$ are represented as higgsino-like $\tilde\chi_1^0$ and they are coded with red colour. In the left panel, the solid (dashed) lines indicate the upper limits coming from current (future) SI direct detection experiments. The black, brown and purple solid lines show XENON1T \cite{Aprile:2018dbl}, PandaX-II \cite{Cui:2017nnn} and LUX \cite{Akerib:2016vxi} upper limits for the  SI  ${\tilde\chi}_1^0$ -  n cross section, respectively, while the green and blue dashed lines illustrate the prospects of the XENONnT and DARWIN for future experiments \cite{Aalbers:2016jon}, respectively. As seen from this panel, all our points are presently consistent with all experimental constraints yet certain DM solutions can  be probed by the  next generation of  experiments. In the right panel, the black, green and purple solid lines show XENON1T \cite{Aprile:2019dbj}, PandaX-II \cite{Xia:2018qgs} and LUX \cite{Akerib:2017kat} upper limits for the SD ${\tilde\chi}_1^0 $ - n cross section, respectively. As seen from this plot, all solutions are consistent with current experimental results, for both singlino- and higgsino-like DM.

\section{Summary and Conclusion}
\label{sec:conclusion}

In this paper, we have explored the low scale and DM implications of an  $E_6$  based UMSSM, with generic mixing between the two ensuing Abelian groups,  
mapped in terms of the standard angle $\theta_{E_6}$. Within this scenario, we have restricted the parameter space such that the LSP is always the lightest neutralino ${\tilde\chi}_1^0$, thus serving as the DM candidate. We have then applied all current collider and DM bounds onto the parameter space of this construct, including a 
refined treatment of $Z'$ mass and coupling limits from LHC direct searches via $pp\to ll$ and $pp\to WW$ processes, allowing for interference effects between their $Z'$ and $\gamma,Z$ components. We have done so as compliance of such a generic $E_6$  inspired UMSSM with all other experimental constraints necessarily requires a gauge kinetic mixing between the $Z$ and $Z'$ states (predicted from RGE evolution from the GUT to the EW scale), which in turn onsets a significant $Z'WW$ coupling.
So that, for $Z'$ masses  in the TeV range, the $Z'\to WW$ decay channel overwhelm the $Z'\to ll$ one, thus producing a wide (yet, still perturbative) $Z'$ state and so that it is the former and not the latter search channel that sets the limit on $M_{Z'}$, at 4 TeV,  significantly  below what would be obtained in a NWA treatment of the $Z'$. To achieve this large $ Z' $ width scenario, the fundamental parameters responsible for it, i.e., the gauge kinetic mixing coefficient and the aforementioned $E_6$ mixing angle, are found to be $ 0.2 \lesssim \kappa \lesssim 0.4 $ and $ -1 \lesssim \theta_{E6} \lesssim -0.8 $ radians, respectively. Curiously, the values of $\theta_{E_6}$ that survive our analysis are not those of currently studied models,
known as   $\psi, N, I, S, \chi$ and $\eta$ types. As for the DM sector,  solutions consistent with all current experimental bounds coming from relic density and direct detection experiments 
were found for two specific LSP compositions: a higgsino-like LSP neutralino with $ 0.9$ TeV $ \lesssim m_{\chi_{1}^{0}} \lesssim 1.2 $ TeV and a singlino-like LSP neutralino with $ 0.9$ TeV $ \lesssim m_{\chi_{1}^{0}} \lesssim 1.6$ TeV. In this respect, we have been able to identify chargino-neutralino coannihilation and  $A$ (the pseudoscalar Higgs state) mediated resonant annihilation as the main channels rendering our DM scenario consistent with WMAP and Planck measurements, with the LSP state being more predominantly singlino-like than higgsino-like. Further, as for SI and SD 
${\tilde\chi_1^0}$ - n scattering cross section bounds from DM direct detection experiments, we have seen that both DM scenarios are currently viable (i.e., compliant with present limits) yet they could be detected by the next generation of such experiments (though we did not dwell on how the two different DM compositions could  be separated herein). In fact, other than in the DM sector, further evidence of the emerging $E_6$ scenario may be found also in collider experiments, in both the $Z'$ and SUSY sectors. In the former case, in the light of the above discussion, it is clear that direct searches at the LHC Run 3 for 
heavy neutral resonances in $WW$ final states may yield evidence of the $Z'$ state, though such experimental analyses should be adapted to the case of a wide resonance. In the latter case, since our  set up yields a rather heavy sparticle spectrum for third generation sfermions ($ m_{\tilde{t},\tilde{b}} \gtrsim 4 $ TeV  and $ m_{\tilde{\tau}} \gtrsim 5 $ TeV) as well as the gluino ($ m_{\tilde{g}} \gtrsim 4 $ TeV),  chances of detection may stem solely from the EW -ino sector, where some relevant masses can be around or just below the 1 TeV ballpark, with   $ pp \to \tilde{\chi}_{2}^{0} \tilde{\chi}_{3}^{0} \to (h/Z)(h/Z) \tilde{\chi}_{1}^{0} \tilde{\chi}_{1}^{0}$   being a potential discovery channel at the HL-LHC. Addressing quantitatively these three future probes of our $E_6$  based UMSSM was beyond the scope of this paper, but this will be the subject of forthcoming publications.

\section*{Acknowledgments}
SM is supported  in part through the NExT Institute and the STFC consolidated Grant No. ST/L000296/1. The work of MF and \"{O}\"{O} has been partly supported by NSERC through grant number SAP105354, and by a grant from MITACS corporation.  The work of YH is supported by The Scientific and Technological Research Council of Turkey (TUBITAK) in the framework of  2219-International Postdoctoral Research Fellowship Program. The authors also acknowledge the use of the IRIDIS High Performance Computing Facility, and associated support services at the University of Southampton, in the completion of this work. \"{O}\"{O}  thanks  the  University  of Southampton, where part of this work was completed, for their hospitality.
\bibliography{Wide_Z_prime}

\end{document}